# HIGH ENERGY APPROACHES TO LOW ENERGY PHENOMENA IN ASTROPHYSICS

SAM M. AUSTIN

*NSCl and Department of Physics and Astronomy*
*Michigan State University, East Lansing MI 48824 USA*
*E-mail: Austin@nscl.msu.edu*

Studies in nuclear astrophysics have long been associated with long runs at small accelerators, measuring ever-decreasing cross sections as one approached (but rarely reached) the energy of reactions in stars. But in recent years pioneering studies have shown that studies at high-energy accelerators can often yield the same information, and in some important cases, provide information not otherwise available. This is particularly so for studies of the properties and reactions of the short-lived radioactive nuclei that play a crucial role in explosive phenomena such as novae, supernovae, and neutron stars. I'll give an overview of some of the possibilities, and then concentrate on two extended examples: measurements of the rates of radioactive capture reactions using Coulomb breakup reactions, and the relationship of charge exchange cross sections and beta-decay strength for L = 1 transitions.

## 1 Introduction

It is a common perception that experimental nuclear astrophysics involves long measurements of small cross sections at lower and lower energies, so as to permit a reliable extrapolation to actual astrophysical energies. This perception is only partially correct. Recent developments, especially of radioactive beams, often permit one to obtain equivalent information with higher energy beams. The high energy experiments commonly yield higher event rates and sometimes yield information not available in the classical approach.

An important advantage is that one can confidently reduce the overall uncertainty in the determination of an astrophysical reaction rate by combining various approaches, with assurance that the systematic uncertainties are independent. For example, if a reaction rate is needed to 5%, four independent approaches at the 10% level will be sufficient, and may lead to a more reliable overall result.

### 1.1 Energy scales

The energy scale for nuclear reactions is given by the location of the Gamow Peak that is determined by the product of the Maxwell Boltzmann distribution of particle velocities at a given temperature and the rapidly increasing barrier penetrability leading to a nuclear reaction. For reactions in the sun, the Gamow-peak energies are in the 5-25 keV range.

*1.2 Resonant and non-resonant processes.*

If a resonance occurs in the Gamow Peak it typically dominates the total reaction rate. For a (p, γ) reaction the rate is $\propto [\Gamma_p\Gamma_\gamma/(\Gamma_p + \Gamma_\gamma)]\exp(-E_r/kT)$; it is only necessary to measure the relevant nuclear widths and the energy of the resonance to specify the rate. If the reaction is non-resonant it is characterized by the S factor, $S = \sigma E \exp(bE^{-1/2})$ and one needs to determine S in the region of the Gamow Peak.

High-energy reactions have commonly been applied to determination of resonant, rates and particularly for the measurement of resonance energies $E_r$. But it is only recently, with the development of radioactive beams, that it has been possible to obtain non-resonant rates using high-energy techniques. In this paper I'll describe a selection of high-energy approaches, some briefly with selected references to the recent literature, and two others in more detail.

**2 Some approaches, briefly**

*2.1 Measurements of the Asymptotic Normalization Coefficient (ANC)*

At low energies radiative capture reactions such as (p, γ) and (α, γ) are dominated by processes occurring far outside the nuclear radius. The cross section then depends on the square of a wave function whose radial dependence, governed by the Coulomb force, is given by the Whittaker function, and whose normalization is given by the ANC[1,2]; the value of the ANC is then sufficient to specify the value of the S factor at E = 0. The most active program is at Texas A&M University where ANCs have been measured by studying transfer reactions at low energies. S-factors have been obtained, for example, for the $^{13}$C(p, γ)$^{14}$N, $^{16}$O(p, γ) $^{17}$F, and $^{7}$Be(p, γ)$^{8}$B reactions. The $^{7}$Be(p, γ)$^{8}$B reaction produces the high-energy neutrinos that dominate the response of solar neutrino detectors. The transfer reactions $^{10}$B($^{7}$Be,$^{8}$B)$^{9}$Be and $^{14}$N($^{7}$Be,$^{8}$B)$^{13}$C were studied [3] to determine the S factor, $S_{17}$, that describes the rate of this reaction. The results are given in Fig.5.

The major issue in this approach is the uncertainty in the optical model potentials (OMPs) necessary to carry out the necessary DWBA calculations. A systematic effort has been made to determine the relevant OMPs with the result that overall precisions of about 10% are possible. The ANC results for the $^{16}$O(p,γ) reaction agree with the direct observation at this level.

## 2.2 The Trojan-Horse Method

The Trojan-Horse (THM) method [4,5] can be regarded an adaptation of the quasi-free knockout method that has been applied in the past to study the momentum distribution of nucleons in nuclei using (p,p') and ($\alpha,\alpha$') reactions. The essence of the THM approach can be obtained from an examination of an example case: study of the low energy cross section of the $^7$Li(p, $\alpha$)$\alpha$ reaction using the $^2$H($^7$Li, $\alpha\alpha$)n reaction. This is diagrammed in Fig. 1. The reaction is induced by $^7$Li ions with an energy above the Coulomb barrier, and the kinematics of the detected $\alpha$'s are chosen so the neutron is essentially a spectator to the reaction between $^7$Li and the proton that occurs at the circled vertex. The Fermi energy of the proton in $^2$H can (partially) compensate for the energy of the $^7$Li, so that low relative $^7$Li-p energies can be studied.

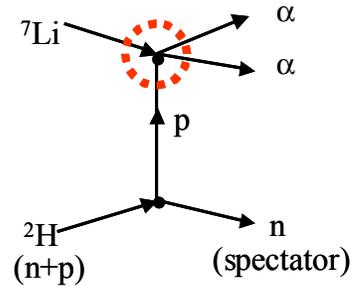

Fig. 1: Trojan-Horse Method for the $^7$Li (p,$\alpha$)$\alpha$ reaction.

Results obtained [6] for the reaction diagrammed in Fig. 1 are shown in Fig. 2. Because of the relatively simple theory employed to connect the three-body and two-body cross sections, absolute cross sections are unreliable; it is necessary to normalize to the directly measured cross sections at higher energies. At lower energies the THM cross section deviates from the direct measurements, presumably because of electron screening corrections: the low-E reactions take place at such large distances from the nucleus that the nuclear charge is somewhat shielded by atomic electrons and the Coulomb barrier is reduced. These screening effects are not present in the THM cross sections. If one accepts the extrapolation of the THM cross section to low energies, then the difference can be ascribed to screening effects and compared to established theories. The surprise is that the effects are much larger than anticipated. In this and the other case studied in detail (6Li(d,αα) for $^2$H(α, $^6$Li), Ref. [7]) the effective screening potential is about twice that expected.

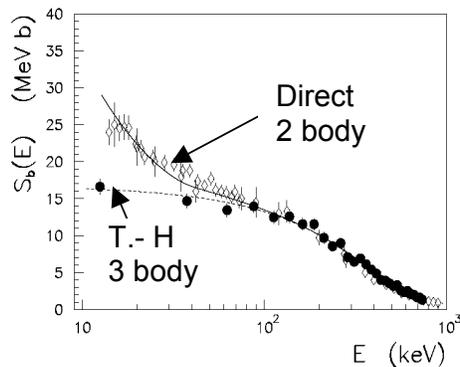

Fig. 2: Direct and Trojan Horse Method S-factors for the $^7$L(p,$\alpha$)$\alpha$ reaction. From Ref. [6].

These studies are in an early state, and their accuracy is not yet clear. But the TH technique has promise for shedding light on screening effects [8]. It may also be possible to extend the technique to radiative capture reactions.

### 2.3 Gamow-Teller strength

It is now well established that Hadronic charge exchange (CEX) reactions such as (p,n) or (n,p) provide an accurate measure of GT strength B(GT), especially for strong transitions [9,10]. Taking advantage of this capability, (n,p) reactions on nuclei in the iron range have been used to measure the strength of electron capture reactions expected to be important in describing supernova core collapse. Recent shell model calculations [11] are in reasonable agreement with these results, but higher resolution data and data on odd-odd and higher mass nuclei, some radioactive, are necessary to establish a truly reliable experimental database. And calculations of supernova evolution using the new theoretical rates show that radioactive nuclei also play an important role in supernovae[12]. As a result of all these considerations, it seems that additional measurements are required, and that they will have to employ CEX reactions involving heavier projectiles.

This raises an important issue. These heavier projectiles are often strongly absorbed and strong absorption means that a limited region of space contributes to the cross section. One would anticipate that by the uncertainty principle, a large range of momenta in the transition form factor would contribute to the CEX cross section. In contrast, B(GT) depends only on the form factor at q≈0. In the cases that have been investigated, however, it has been found that even strongly absorbed projectiles (as in the ($^{12}$C,$^{12}$N) reaction [13]) yield accurate B(GT). Osterfeld *et al.* [14] have shown that this surprising result follows from two facts: that the CEX reaction is only sensitive to the form factor over a limited (although fairly large) range of q, and that, (in the cases studied) the transition form factors have essentially the same q dependence in this region.

This conclusion makes it possible to use ($^3$He,t), (t, $^3$He), and ($^7$Li,$^7$Be) reactions on stable nuclei to obtain $GT_+$ and $GT_-$ strengths with high resolution; to use the ($^6$Li,$^6$Li$^*$ ($0^+$ T=1)) reaction to measure $GT_0$ strength by inelastic scattering; and to use some of these reactions in inverse kinematics to study GT strength for radioactive nuclei using radioactive beams. It is not yet clear what techniques will prove most useful for these inverse kinematics studies, because the outgoing light particles have such low energies. But at a minimum, low-energy-resolution data will be obtainable by catching and identifying the heavy product in high-resolution spectrometers such as the NSCL S800.

**3  Coulomb dissociation-a detailed example**

Only recently has Coulomb dissociation (CD) been employed to study the breakup of radioactive nuclei of interest for astrophysics and thereby to obtain information about the related radiative capture process. The principle is simple and we illustrate it by considering the CD of $^8$B and how it can be used to determine the S-factor, $S_{17}$, for the $^7$Be(p, $\gamma$)$^8$B reaction. Upon passing a high-Z nucleus a fast $^8$B projectile absorbs a virtual photon, and breaks up into $^7$Be and a proton. The energy of the absorbed photon is determined by measuring the momenta of the products and the cross section for the (p,$\gamma$) reaction is determined by detailed balance. The event rate is larger for CD than for (p,$\gamma$) because thick targets can be used and because the momentum ratio in detailed balance favors CD. There is however a complexity: detailed balance applies only to gamma rays of a given multipolarity, but the contributions of various multipoles differ for the two processes. For $^8$B, L=1 photons dominate both processes, but E2 photons, entirely negligible for (p,$\gamma$), can contribute significantly to CD. This was recognized early, but still remains a source of significant controversy and some uncertainty.

The first experiments involving radioactive nuclei were those of Motobayashi, *et al.* [15] at RIKEN, on the breakup of $^{14}$O and of $^8$B[16]. More recently measurements on $^8$B have been done at higher energies at GSI[17] and $^8$B[18,19,20] and $^9$Li [21] at MSU/NSCL. The results for $^{14}$O have been confirmed by (p,$\gamma$) experiments and other CD measurements. The results for $^8$B are also in general agreement, as we shall see in Fig. 5.

*3.1  Inclusive Experiments*

Because of the concern about E2 contributions for $^8$B, all experiments have attempted to assess its contributions. The RIKEN and GSI experiments obtained limits on E2 values that were significantly smaller than theoretical estimates (see summary in [20]). At MSU/NSCL, an attempt was made to obtain a more accurate value by observing the interference between E1 and E2 amplitudes in the inclusive distribution of $^7$Be longitudinal momenta; the $^7$Be ions were observed in the high acceptance S800 spectrograph. The results, shown in Fig. 3, are strongly asymmetric; such an asymmetry can occur only if E2 amplitudes contribute. For B(E2) = 0, the distribution would be symmetric in first order perturbation theory (PT) while higher order effects would produce the opposite asymmetry.

We have analyzed these data in PT and obtain a value of E2 strength about half that predicted by theoretical estimates; if higher order effects were included in the analysis, one would expect to obtain a somewhat larger value [22]. We have checked this inclusive result by examining the angular distribution of protons in the c.m. of $^8$B as observed in the exclusive experiment [20]. It is consistent with the inclusive result, but with lower statistical precision. The reason for the disagreement with the earlier data is not understood, but may possibly result from the contributions of nuclear processes in the RIKEN and GSI experiments.

### 3.2 Exclusive measurement

Exclusive studies of $^8$B breakup were carried out at 83 MeV/nucleon at MSU/NSCL. The breakup products were separated in a magnetic field and detected and identified by an array of multiwire drift chambers and plastic scintillators [20]. The results shown in Fig. 4 are for a scattering angle cutoff of 1.77 deg, or a classical impact parameter around 30 fm. First order PT calculations show that the E2 contribution dominates at low $E_{rel}$, but is less that 10% above 130 keV. The Continuum Discretized Coupled Channel (CDCC) calculations show that nuclear contributions are small. After small corrections for E2 effects and the contributions of breakup leading to $^7$Li (478 keV), we obtain $S_{17}$ = 17.8 (+1.4/-1.2) keV b. This value is consistent with most other recent results with the single exception of the recent and as yet unpublished result of Junghans, *et al.* [26]. See Fig. 5. for details and a weighted mean.

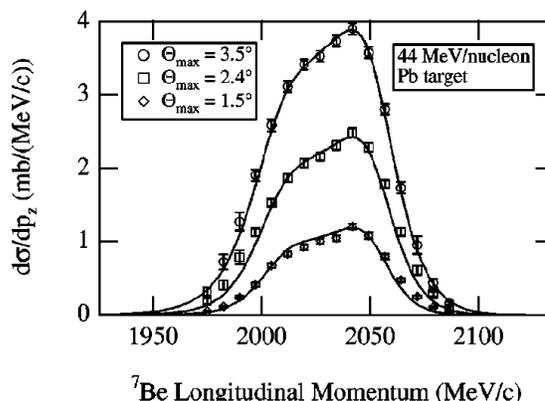

Fig. 3: Measured longitudinal momentum distributions of $^7$Be fragments from the CD of 44 MeV/nucleon $^8$B on Pb; several maximum $^7$Be angle cuts are shown. Also shown are first-order perturbation theory calculations convoluted with the experimental resolution (5 MeV/c). From [18, 20]. See the text for details.

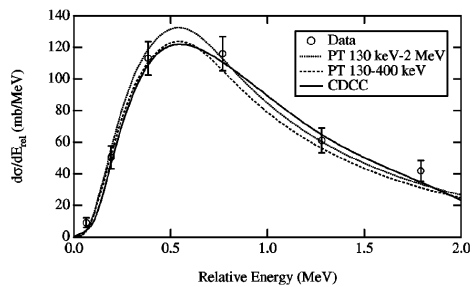

The result of Junghans, *et al.* [26] has not been included in the weighted average, because it is inconsistent with most recent results. Yet [26] describes a carefully done experiment that must be taken very seriously. A major goal of future work to understand this difference. It is perhaps worth noting that both the CD and ANC methods have been checked against particular direct measurements to good accuracy. In the case of the MSU/NSCL measurements care was taken to choose the conditions of observation to minimize the confounding effects (E2, nuclear) of the CD method.

Fig. 4: Measured cross sections for breakup of 83 MeV/nucleon $^8$B ions with $^8$B scattering angles of ≤ 1.77 deg. PT calculations normalized to the data over the regions noted, and CDCC calculations are also shown. The lowest energy point was excluded from the fit because E2 contributions are large at low relative energy. From [19,20].

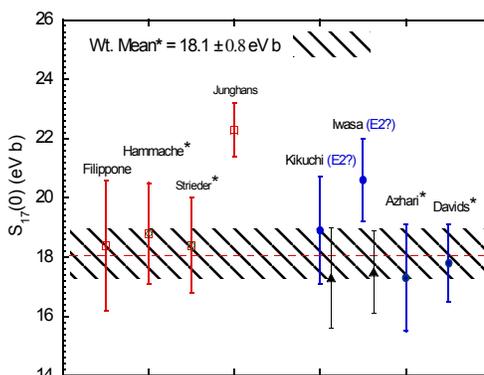

Fig. 5: Results for $S_{17}$. The results to the left of the horizontal break are the direct results, while those to the right have been measured by CD or from the ANC (See refs [23, 24, 25, 26, 16, 17, 3, 20] in horizontal order, left to right). The cross hatched area is the weighted mean of the results marked with *. The CD results of Kikuchi *et al.* and Iwasa, *et al.* have been excluded because of the present author's concern about E2 contributions. If they are included the weighted mean is 18.7 eV b. Earlier work is reviewed in [27].

Another issue involves possible E2 contributions in the RIKEN and GSI measurements. We believe that the MSU measurements of the E2 effects are the most sensitive, because they observe interference phenomena. For that reason, we have made a rough estimate of the effects of E2 amplitudes on the GSI/RIKEN results. The estimated changes are shown by the triangles near the bottom of the cross hatched area.

### 4  Can charge exchange reaction measure L=1 electroweak strength?

It appears that forbidden electroweak transitions can play an important role in astrophysics. Neutrinos can excite the spin-dipole resonance in abundant alpha-particle nuclei; part of the time the excited nucleus will be unbound and can emit nucleons, leading to the production of rare nuclides such as $^{7}$Li, $^{11}$B, and $^{19}$F. Similar phenomena can modify the distribution of nuclides formed in the r-process. Neutrino reactions also play a role in the transport of energy by neutrinos in supernova explosions and are the basis for the observation of supernova neutrinos in terrestrial detectors. Yet direct observation of neutrino induced reactions will be experimentally feasible in only a very few cases.

In this circumstance, it will be necessary to rely on hadronic reactions to measure the required strengths, following the approach used for determination of allowed (GT) strength that was discussed in section 3.3. It is not known whether this procedure will work reliably, for the reasons discussed in section 2.3 and because the qualitative conditions are still less favorable: larger momentum transfers and contributions of tensor forces will be a confounding factor. We [28] have investigated two questions. (1) Are σ (CEX) and B(L=1) proportional when both are calculated with the same wave functions. A positive answer to this questions is a necessary condition for CEX to be useful. (2) What range of momentum transfer is important in the transition form factor F(q').

We considered a simple case: $^{12}$C(p, n)$^{12}$N at $E_p$ = 135 MeV [29]. A simple eikonal model, taking into account the real and imaginary parts of the OMP, was used to provide insight into the reaction mechanism. This permitted the evaluation of a sensitivity function S(q, q') which characterizes the range of q' in the transition form factor which contributes at a given asymptotic momentum transfer q. Specifically, $T(q) = T_{PW} + \int dq' S(q,q') F(q')$ Where $T_{PW}$ and $T(q)$ are the plane wave and total (p,n) reaction amplitudes. The approximations involved were checked against the distorted wave impulse approximation; the two predictions were in good agreement with each other and with angular distributions for J=1$^-$ and 2$^-$ states.

The results are shown in Figs. 6.

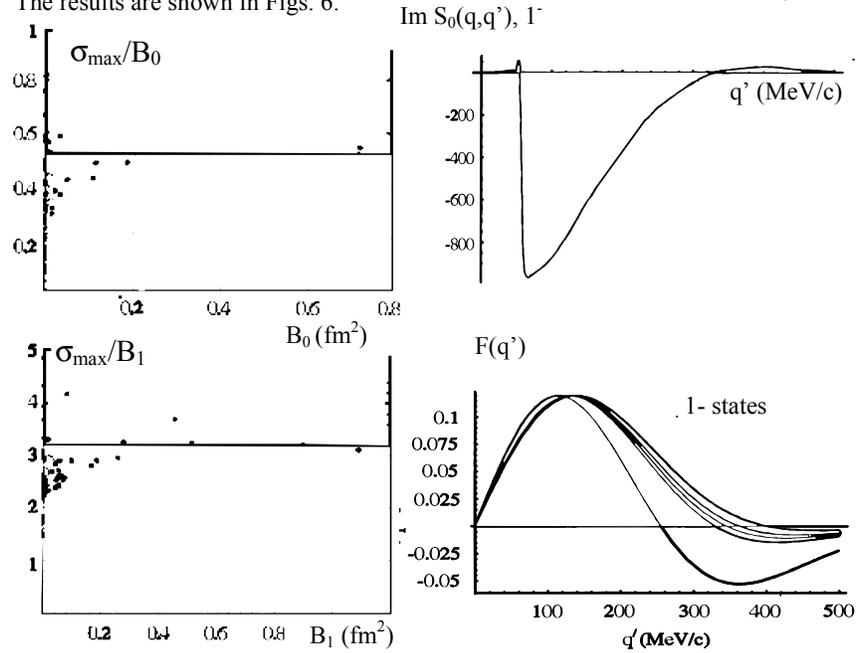

Fig. 6: The left panel shows the ratio of calculated cross sections (at maxima) and calculated B(L=1) for 0$^-$ and 1$^-$ states. The right panel shows the sensitivity functions (imaginary part) for a 1$^-$ state, and the q' dependence of the transition form factors for various 1$^-$ states.

The conclusions we draw from Fig. 6 are twofold. For $B_J > 0.1$ fm$^2$, $B_J \propto \sigma(CEX)$ to within 10-15%, about the same level of proportionality observed for GT transitions. And the location of the peaks in the sensitivity function $S(q,q')$ indicates that the (p,n) reaction amplitude is dominated by q' for which the transition form factors for various states are proportional. This gives us some confidence that CEX reactions can be used to measure the bulk of $L = 1$ electroweak strength. It remains to generalize to other systems: to determine whether $S(q, q')$ is localized for heavier systems and more strongly absorbed articles, and whether the $F_J(q')$ are similar for the important q'.

## 5  Final comments

### 5.1  Techniques discussed

For techniques involving high-energy particles we find:

> In inverse kinematics at high beam energies, thicker targets and large solid angle coverage yield large event rates.
> For low energies, screening effects are absent, an advantage for some reactions
> It remains to reduce the systematic uncertainties that may still be present in these new techniques. But they already approach in accuracy a typical experiment using standard techniques.
> A combination of techniques with independent systematic errors may offer the best chance of obtaining accurate results in difficult cases.

### 5.2  Other measurements

Use of high energy beams may give a larger reach toward the drip line even for experiments that can be done at low energy. For example:

> Identification of beam particles at high energy and then stopping them to measure β-decay half lives and to perform spectroscopy of β-delayed neutrons and gamma rays near the r-process path.
> Mass measurements with moderate (but probably sufficient) accuracy near the r-process and rp-process paths.